\documentclass[12pt,preprint]{aastex}

\shorttitle{UHECRs}
\shortauthors{Ouyed et al.}

\begin{document}

\title{\large Ultra-high energy cosmic rays from hypothetical Quark Novae}

%
\author{{\sc R. Ouyed}}
\affil{Department of Physics and Astronomy
 University of Calgary, 2500 University Drive NW, Calgary, Alberta, T2N 1N4 Canada}
\email{ouyed@phas.ucalgary.ca}

\author{{\sc P.~Ker\"anen}}
\affil{Nordic Institute for Theoretical Physics, 
DK-2100 Copenhagen, Denmark;
Centre for Underground Physics in Pyh\"asalmi,  
P.O.~Box~22, 86801 Pyh\"asalmi, Finland}

\and

\author{{\sc J.~Maalampi}}
\affil{Department of Physics,
P.O.~Box~35, FIN-40014 University of Jyv\"askyl\"a, Finland;
Helsinki Institute of Physics, P.O.~Box~64, FIN-00014 University of Helsinki, Finland}

\begin{abstract}
We explore acceleration of ions in the Quark Nova (QN)
scenario, where a neutron star 
experiences an explosive phase transition into a quark star
(born in the propeller regime).
In this picture, two cosmic ray components are isolated: one related 
to the randomized pulsar wind and the other to the propelled wind, 
both boosted by the
ultra-relativistic Quark Nova shock. The latter component acquires
energies $10^{15}\, {\rm eV}<E<10^{18}\, {\rm eV}$ while the former, 
boosted pulsar wind, achieves ultra-high energies $E> 10^{18.6}$~eV.
The composition is dominated by ions present in the pulsar wind
in the energy range above $10^{18.6}$~eV, while at energies below 
$10^{18}$~eV the propelled ejecta, consisting of the 
 fall-back neutron star crust material 
 from the explosion, is the dominant one.
Added to these two components, the propeller injects 
 relativistic particles with Lorentz factors 
$\Gamma_{\rm prop.} \sim\, 1-1000$, later to be accelerated 
by galactic supernova shocks. 
The QN model appears to be able
to account for the extragalactic cosmic rays
above the ankle and to contribute a few percent
of the galactic cosmic rays below the ankle. 
We predict few hundred ultra-high 
energy cosmic ray events above $10^{19}$~eV for the Pierre 
Auger detector per distant QN, while some thousands are 
predicted for the proposed EUSO and OWL detectors.
\end{abstract}


\section{Introduction}

Large efforts have been devoted to explore the origin of cosmic
rays. Most puzzling are the observed ultra-high energy cosmic ray
events (UHECRs) above the GZK-cutoff \citep{greisen66,zatsepin66}:
protons lose energy drastically due to pion photo-production
processes on the cosmic microwave background (CMB) at energies
higher than $7\times 10^{19}$~eV, and this limits the proton mean
free path to some tens of megaparsecs (see \cite{blasi03}
and \cite{bahcal03} for more discussion
on the GZK cut-off). Two distinctly different
classes of models are commonly considered, the top-down and
bottom-up scenarios. In the former ones, UHECRs are associated
e.g. with the decay of some supermassive particles, whereas in the
latter scenarios UHECRs are assumed to be accelerated by
astrophysical objects (e.g. \cite{sigl94,venkatesan97,biermann98},
and for a latest review we refer the interested
reader to \cite{ostrowski02,arons03,sigl03,olinto04}).
 A growing number of bottom-up explanations
have been proposed, including active galactic nuclei, gamma ray
bursts and neutron stars. However, at present there seems to be no
clear association of UHECR events with any of these objects
although by tuning the model parameters one enables the highest
energy events to be accounted for.

Identifying sources of the
observed UHECR events remains uncertain and debated
leaving room for speculation (\cite{nagano,torres,stanev04}). As such we wish to  
 consider in this paper the possibility that
 hypothetical quark novae (hereafter, QNe;
 \cite{ouyed02}) contribute to the cosmic ray flux, especially 
above $10^{18}$~eV.
In the QN explosion the core of a neutron star (NS) shrinks into the
equilibrated quark object/star (QS). The overlaying crust material free-falls
following the core contraction releasing enough energy
to form an ultra-relativistic ejecta.
The ultra-relativistic shock interaction
with its surroundings environment (namely the randomised
relativistic wind of the
progenitor) leads to the first cosmic ray component in our model.
 The compact remnant acting as a magnetohydrodynamic
propeller provides the second component, injecting ions to be boosted
 by the ultra-relativistic quark nova shock. Most of the propelled wind 
of relativistic particles is injected into interstellar space with
low energies (from GeV to TeV), and these particles can be further on 
accelerated by galactic supernova shocks. Thus the propeller can 
contribute to cosmic ray flux also below the knee as an injector.
It appears that the QN model 
 can largely contribute for the extragalactic cosmic rays
above the ankle and contribute a few percent
of the galactic cosmic rays around and below the knee.

The main assumption in this paper is the conversion of a highly magnetised,
rapidly spinning NS to a QS. This phase conversion
(or second explosion) -- which remains to be confirmed -- 
leads to unique conditions where two UHECR components are
feasible. It is this unique aspect of the model that
makes this investigation worth pursuing.
We note however that the approach presented here  
 (including the acceleration mechanisms as related
to the dynamics of the explosion)
 borrows heavily from what has already
been presented in the literature for models of UHECRs
in the context of isolated pulsars and binary coalescence
 (e.g., \cite{venkatesan97,gallant99,blasi00,arons03} to cite only few).

This paper is presented as follows: We start with a
brief  review of the concept of quark nova and its
features (\S2). In \S3 we describe conditions under which the
compact remnant is born in a propeller regime.
In \S4 we explain how cosmic rays are produced and isolate
the two components (with
energies $10^{15}\, {\rm eV}<E<10^{18}\, {\rm eV}$ and 
 $E> 10^{18.6}$~eV). Here we describe the QN compact
remnant as an injector of relativistic particles to the
galaxy.
 In \S5 we predict
UHECR events in future detectors and discuss multiple
events and the isotropy of sources in our model. A discussion
 and a conclusion follow in \S6.

\section{Quark Nova}

It has been suggested that the core of neutron stars may deconfine
to a composition of up ($u$) and down ($d$) quarks
during or shortly after some supernova explosions when the central
density of the proto-neutron star is high enough to induce phase
conversion (see e.g. \cite{dai95,xu01}).
It has also been speculated that when the density in the core increases
further, a phase with strange quarks ($s$) becomes energetically
favoured over the pure {\sl (u,d)} phase and soon the entire star
is contaminated and converted into this {\sl (u,d,s)} phase. This
is one of the scenarios introduced to convert an entire NS 
 to a QS (e.g., \cite{cheng96,bombaci00}).
 In the QN picture the
{\sl (u,d,s)} core is
assumed to shrink to a corresponding stable quark object before the
contamination is spread over the entire star. By
physically separating from the overlaying material (hadronic
envelope) the core drives the collapse (free-fall) of the
left-out  matter leading to both gravitational energy and phase
transition energy release as
 high as $E_{\rm QN}\simeq 10^{53}$~ergs (\cite{ouyed02,keranen05}).

\subsection{Quark Nova ejecta and compact remnant}

The QN-ejecta consist mainly of heavy nuclei
(the NS crust), protons, electrons and
neutrons.  The total amount of the ejecta has been  
estimated to be of the order of $(0.001 -
0.01)M_{\odot}$ with corresponding Lorentz factor 
 $\Gamma_{\rm QN}= \epsilon~ E_{\rm QN}/M_{\rm ejec.} c^{2} \sim 10\, -
\,100$, where $\epsilon$ is the percentage of the QN energy
transfered to the QN shock front which we assume to 10\%.

The radius of the newly formed QS is given by $R \simeq
R_{\rm NS}(\rho_{\rm NS}/\rho)^{1/3}$ where the NS core
density, $\rho_{\rm NS}$, and the QS density, $\rho$,
scale as $\rho_{\rm NS}/\rho\simeq 0.1-0.2$ \citep{ouyed02,keranen05}
 -- {\it hereafter parameters with no
subscripts refer to the quark star}. The QS
spins up during the phase transition due to contraction, with a 
rotational period 
$P = P_{\rm NS}(R/R_{\rm NS})^{2} P_{\rm NS}(\rho_{\rm NS}/ \rho)^{2/3}\simeq
P_{\rm NS}/4$.
In the process, the magnetic field is amplified, 
$B = B_{\rm  NS}(R_{\rm NS}/R)^{2}
= B_{\rm  NS}(\rho/\rho_{\rm NS})^{2/3}\simeq 4 B_{\rm NS}$.

\subsection{Fall-back material}

Fall-back of material onto a newly formed quark star may take
place (similar to what has been
suggested in the supernova case; \cite{cheval89}). In the early stages the accretion rate is given by 
\begin{equation}
\dot{m} \simeq 10^{28}\, {\rm g/s}\, (\frac{\rho_{\rm ff}}{10^{6}\
{\rm g/cm^3} }) (\frac{R}{10 \, {\rm km}})^{3/2} (\frac{M}{1.5\,
M_{\odot}})^{1/2}\, ,
\label{eq3}
\end{equation}
where
$\rho_{\rm ff}$ is the average density of fall-back matter ($10^6$~g/cm$^3$,
representing the crust material, and the matter 
below the neutron drip line densities).
The hyper-Eddington accretion rate given above is understood
by noting that (i)
the fall-back matter in the initial phase of the explosion is not accreted onto
the surface (but is propelled away before radiating), and (ii) the crust would
not form in the early stages leaving
the quark star bare (not subject to the Eddington limit
since the bulk of the star
is bound via strong interaction rather than gravity; see e.g. \cite{alcock86,zhang00}, and \cite{xu03} for a recent discussion on 
the
matter).

\section{Propeller regime}

For the remainder of this paper, we consider a QS (NS) mass  of
$M=1.5M_{\odot}$ ($M_{\rm NS}\sim 1.5M_{\odot}$), a
radius $R=10$~km ($R_{\rm NS}\sim 12.5$~km), a surface
magnetic field of $4\times 10^{14}$~G ($B_{\rm NS}\sim 10^{14}$
G). Since the fastest pulsar has a period of $1.56$~ms \citep{backer82},
we use $2$~ms as a representative value for the period $P$
of the new born quark star ($P_{\rm NS}\sim 8$ ms).
Clearly these are unique constraints (millisecond,
highly magnetised NS) favouring a scenario where
the QS forms immediately following a SN explosion,
or that where the NS has been spun up
by accretion from a companion.

The newly born quark star is defined by its three critical radii:
the Keplerian ``co-rotation radius''
\begin{equation}
R_{\rm c}= 27\, {\rm km}\, (\frac{M}{1.5\, M_{\odot}})^{1/3}(\frac{P}{2\, {\rm ms}})^{2/3}\, ,
\label{eq4}
\end{equation}
the magnetospheric radius at which the ram pressure
of the in-falling matter balances the magnetic pressure
\begin{eqnarray}
R_{\rm m}&=& \big( \frac{B^2R^6}{2\dot{m}\sqrt{2GM}}
\big)^{2/7}=55\, {\rm km}\,
\big(\frac{B}{4\times 10^{14}\, {\rm G}}\big)^{4/7}
\times \nonumber \\
&\times &\big(\frac{10^{28}\,
{\rm g/s}}{\dot{m}}\big)^{2/7} \big(\frac{R}{10\, {\rm
km}}\big)^{12/7}\big(\frac{1.5\, M_{\odot}}{M}\big)^{1/7}\, ,
\label{eq5}
\end{eqnarray}
(see, e.g. \cite{frank92}),
and the light cylinder radius
\begin{equation}
R_{\rm lc} = \frac{c}{\Omega} = 96\, {\rm km}\, (\frac{P}{2\, {\rm ms}})\, .
\label{eq6}
\end{equation}
Given our fiducial values, the QS is born in the propeller
regime \citep{shvartsman70,illarionov75}, i.e.
$R_{\rm c} < R_{\rm m} < R_{\rm lc}$, where the infalling
material may be accelerated in a wind that carries away
angular momentum from the magnetosphere and hence from the QS
itself (\S4.2).
Note that the propeller regime is only
possible if the quark star is born with
a magnetic field exceeding the
critical value $B_{\rm c}\ge 10^{14}$ G,
which corresponds to a parent neutron star
initial magnetic field  $B_{NS,c} = B_{\rm c}/4 \ge 2.5\times 10^{13}$ G.
 Here ``c" stands for ``critical" while the factor 4 (derived from our fiducial
values) is the result of the amplification
of the magnetic field following the quark-nova explosion.
These critical values explain our assumption
of a highly magnetised, rapidly spinning NS.

\subsection{Propeller lifetime}

The star loses rotational energy at a rate
defined by the gravitational radiation losses ($\dot{E}_{\rm grav.}$),
electromagnetic radiation losses ($\dot{E}_{\rm em}$) and the propeller's
torque ($\dot{E}_{\rm prop.}$). That is,
\begin{equation}
\frac{{\rm d}\Omega}{{\rm d}t} \equiv \dot{\Omega} -\frac{\dot{E}_{\rm prop.}+\dot{E}_{\rm em}+\dot{E}_{\rm
grav.}}{I\Omega}
\label{eq7}
\end{equation}
with \citep{shapiro83}
\begin{eqnarray}
-\dot{E}_{\rm grav.}&=&\frac{9}{5}\frac{GI^2\epsilon^2\Omega^6}{c^5}
=6.8\times10^{47}{\rm erg}/{\rm s}\, \times\nonumber\\
&\times&\big(\frac{2 \, {\rm ms}}{P} \big)^6 \big(\frac{I}{10^{45}\, {\rm g \,
cm}^2} \big)^2 \big(\frac{\epsilon}{0.01}\big)^2\, ,
\label{eq8}
\end{eqnarray}
where $I$ is the moment of inertia
and $\epsilon$ the equatorial eccentricity; and with \citep{manchester77}
\begin{eqnarray}
-\dot{E}_{\rm em}&=&\frac{4B^2R^6\Omega^4}{9c^3}2.6\times 10^{46}{\rm erg}/{\rm s}\times\nonumber\\
&\times&\big(\frac{2 \, {\rm ms}}{P} \big)^4 \big(\frac{B}{4\times 10^{14}\,
{\rm G}} \big)^2 \big(\frac{R}{10\, {\rm km}}\big)^6\, .
\label{eq9}
\end{eqnarray}
The spin-down due to the propeller is \citep{menou99}
\begin{equation}
-\dot{E}_{\rm prop.}=2\dot{m}c^2=1.8\times 10^{49} {\rm erg/s}\
\big(\frac{\dot{m}}{10^{28}\, {\rm g/s}}\big) \, .
\label{eq10}
\end{equation}
The expression above assumes that the material flung away by the
propeller effect has been accelerated to an angular speed
corresponding to that of the star.
We note that
gravitational losses would be important in the spin-down for very short
periods $P<2$~ms (not our case) and propeller losses dominate
over the electromagnetic dipole radiation losses given our
accretion rates.

The propeller regime lifetime can be obtained using
Eq.~(\ref{eq7}) and Eq.~(\ref{eq10}). We find, assuming a constant
accretion rate,
\begin{equation}
 t_{\rm prop.} \simeq 10^3\ {\rm s}\ \big(\frac{I}{10^{45}\,
 {\rm g\, cm^2}}\big) \big(\frac{10^{28}\, {\rm g\ s^{-1}}}{\dot{m}}\big)
(\frac{1}{P_{\rm i}^2}-\frac{1}{P_{\rm f}^2})\ ,
\label{eq11}
\end{equation}
where $P_{\rm i}$ and $P_{\rm f}$, initial
and final periods, are in milliseconds.
For a constant accretion rate, and taken the total amount of the QN
ejecta, the lifetime of the propeller phase would not exceed a
hundred seconds during which time
the QS would have spun-down by no more than 30\% thus remaining within
milliseconds period.
Following this phase, the spin-down is governed by the magnetic
dipole radiation losses of the QS.

\section{Production of the two cosmic ray components}

\subsection{Acceleration of the pulsar wind}

We assume that the pulsar wind is
composed of electron-positron pairs and ions, as would be expected
if pulsars operate as open-circuited systems with ions
carrying the return current \citep{hoshino92}. Some
fraction of ions can be heavy like iron, originating
from the surface of the pulsar. We also assume that the wind
has been subject to a termination shock as to acquire a randomised
direction of particle momenta \citep{gallant99}.
The number density of electron-positron pairs is
$n_{\rm GJ}=B_{\rm NS}(r) \Omega_{\rm NS}/(4\pi ec)$
($n_{\rm GJ}$ is the Goldreich-Julian density; \cite{gj69}) where
$e$ is the electron charge. 

We adopt $\gamma_{\rm w}\sim 10^6$ as the relativistic factor
of most of the wind particles (see e.g. \cite{gallant94}),
and the ratio of ions to electron-positron pairs to be
$\alpha=n_{\rm i}/n_{\rm GJ} \sim 10^{-3}$; ions carrying
most of the energy of the wind.
For future purposes, we write the ion density as
\begin{eqnarray}
n_{\rm i}(r)= \alpha\, n_{\rm GJ}\, \big(\frac{ R_{\rm NS} }{r}
\big)^\beta \, ,
\label{ni}
\end{eqnarray}
where the index $\beta$ describes the radial dependency.

The QN ejecta will interact with the parent NS wind.
The randomised wind particles
will be boosted with one shock crossing by a factor of $2\Gamma_{\rm QN}^2\sim
2\times 10^{4}$ before they leave the region\footnote{Particles do not
have time to re-isotropise upstream before being overtaken by the
QN shock; the acceleration consists of only one cycle.}.
The maximum number of particles that can be
accelerated is limited by the QN fireball energy and can be written
as (here the energy of the pairs will be negligible, since
energetically they are at much lower gamma factors due to synchrotron
losses)
\begin{eqnarray}
N_{\rm wind}\sim \frac{\epsilon~E_{\rm QN}}{ 2\Gamma_{\rm QN}^2 \gamma_{\rm w} Z m_{\rm p}
  c^2} \simeq \frac{3\epsilon}{Z}\times 10^{45}  \, ,
\label{wind}
\end{eqnarray}
where $m_{\rm p}$ is the proton mass, $Z$ can vary from $1$ to $26$ (iron) 
and $\epsilon$ is the fraction of QN energy transferred via
the QN shock to the pulsar wind particles .
 If all ions are protons
and approximately accelerated to the energies of the ankle
(around $10^{18.8}$~eV), then up to $10^{45}$ particles would
gain that energy. Heavier ions will correspondingly reach higher
energies with the same total relativistic factor
$2\Gamma_{\rm QN}^2\gamma_{\rm w}$, e.g. iron will easily gain energies over
$10^{20}$~eV. Therefore in this model at higher energies there 
should be a natural increase in heavier ions.
Since the above follows closely calculations already
presented in \cite{gallant99}, the corresponding spectrum also in our case 
can be shown to be ${\rm d}N/{\rm d}E\sim E^{-2}$ or even flatter given
its dependency on $\Gamma_{\rm QN}$.
We would like to emphasize
that further studies of our model are needed in order
to make better predictions of the spectrum.

Such an injection spectrum might agree with
 AGASA measurements but let us 
 recall that in the high energy region we discuss 
here, currently there seems to be a disagreement between
the AGASA ground array (Takeda et al. 1999)
 and the HiRes fluorescence detector (Abu-Zayyad et al. 2002)
which seems consistent with the GZK-cutoff.
Clearly there is a need for much larger experiments
 such as Auger, EUSO, and OWL, that can increase the number of detected
events by one or two orders of magnitude before the injection
spectrum is known conclusively. As we have said, for now, our model
lacks the details to predict the exact spectrum.

The number of particles per
unit volume from QNe in all the galaxies is $J^{\rm QN}=(c/4\pi)N_{\rm
wind}n_g T_{\rm loss}(E_a)\nu_{\rm QN}$, where $T_{\rm loss}(E_a)$ is the
residence time of a particle at the ankle initially injected at
higher energy \citep{arons03}.  The QN rate per galaxy
is given by $\nu_{\rm QN}$ while $n_g$ is the galaxy density.
With $\nu_{\rm QN}=10^{-6}$~yr$^{-1}$ and
 $n_g=0.02$~Mpc$^{-3}$,
this implies
\begin{equation}
J^{\rm QN}_{1} (E> 10^{18.8}\, {\rm eV})\sim
\frac{5.31\epsilon}{Z}\times 10^{-18}\, {\rm cm}^{-2}~{\rm s}^{-1}~{\rm ster}^{-1}\, ,
\label{eq15}
\end{equation}
as compared to the observed value $J^{\rm obs}(E>10^{18.8}\ {\rm eV}) \sim 3\times 10^{-18}$~cm$^{-2}$~s$^{-1}$~ster$^{-1}$ (1
event per square kilometer per year; \cite{lawrence91,takeda98}), assuming $Z=1$. 
For particles at energies higher than $10^{18.8}$ eV
we  used an average $T_{\rm loss} \sim 1$~Gyr (Biermann\&Strittmatter 1987;
 Berezinsky \& Grigoreva 1988; Bertone et al. 2002;
see also Figure 5 in Sigl 2004b).

The QN flux calculated above is short of the observed
one  unless the QN is very efficient in transferring
  energy via the QN shock into relativistic particles.
However the presence of large scale magnetic field complicates the
 interpretation of UHECRs data
and could lead to  an overestimate of the observed flux.
More specifically, the energy spectrum
of particles emitted from a (nearby) source depends very strongly
 on the  relative position of the observed source and the observer
 to the direction of the large scale field. In some
cases the flux can be enhanced by a factor of $\sim 100$
for energies below $10^{19}$ eV (see Stanev et al. 2003
and references therein).
  The observed flux and spectrum might not reflect the ones at the source
making uncertain a direct comparison with the numbers derived above.

\subsubsection{Acceleration timescale}

The ejecta remain in expansion phase until the shock starts to
decelerate.
The transition between these
two phases occurs at a radius $R_{\rm QN, d}$
at which the energy in the
swept-up material becomes of the order of the energy released in the
QN, i.e when a large fraction of the fireball energy has been used
to reaccelerate the wind particles (\cite{blandford76,gallant99}). Using Eq.(\ref{ni}),
\begin{eqnarray}
R_{\rm QN, d} &\simeq&
\left( \frac{ E_{\rm QN} (3-\beta) }{ 4\pi\alpha n_{\rm GJ}
R_{\rm QN}^\beta  2\Gamma^2\gamma_{\rm w}Z m_{\rm p} c^2 }
\right)^{1/(3-\beta)} \, ,
\label{RQNd}
\end{eqnarray}
which gives
$R_{\rm QN,d}\sim 5.5\times10^{13}/\sqrt{Z}$~cm for $\beta=1$.
If the ion density decreases as $\sim 1/r^2$ ($\beta = 2$),
the fireball will
be in the expansion phase much longer and the distance should
be defined by using interstellar matter density together with
the wind (we estimate $R_{\rm QN,d}\ge 10^{15}$ cm for $\beta =2$).
These numbers will define the timescale of the highest energy cosmic
ray component acceleration.
The case with $\beta\sim1$ would more closely follow the magnetic field
value and therefore the charge density of the pairs. We use
it as an example. The corresponding time is
$t_{\rm QN,d}=R_{\rm QN,d}/c\sim 1700/\sqrt{Z}$~s. If all wind particles
are iron, then the timescale would be shorter, $330$~s.
One should notice however that
the ion density profile in the pulsar wind bubble is not known and
therefore the timescale of the expansion phase could be much
larger (with $\beta\sim2$ up to weeks). 

\subsection{Acceleration of the propelled wind}

The second cosmic ray component occurs when rapidly rotating millisecond
pulsars with large magnetic fields, $B > 10^{14}$ G, undergo
a QN explosion. In this case the 
born QS is in the propeller regime (see \S 3).
The material flung away by the propeller expands as a
magneto-hydrodynamic wind to reach the speed of light (with
Lorentz factors up to $1000$ at the
light cylinder for $B>10^{14}$ G; 
see \cite{fendt04} for detailed calculations)\footnote{
\cite{shvartsman70} first suggested that relativistic
particles can be formed at the propeller stage by a rapidly
rotating magnetic field (see also \cite{kundt90}).}.
The MHD nature of the propeller ensures that acceleration occurs
at large distances from the star where synchrotron losses
are likely to be minimum.
As demonstrated in \cite{fendt04} the magnetic field
above the Alfv\'en surface is predominantly toroidal. Such
a geometry will allow the particles to escape freely along the
poloidal direction in the acceleration zone without being
deflected by the magnetic field lines.
Particles with $\Gamma_{\rm prop.} >
\Gamma_{QN}$ can reach eventually the QN shock, as to reach Lorentz
factors as high as $2\Gamma_{\rm QN}^2\times\Gamma_{\rm prop}
\simeq 2\times10^7$.
It is roughly $10^{16}$~eV for protons and $10^{18}$~eV for iron in one
shock crossing. 
Since an efficient propeller needs strong magnetic field and a short period, 
this component is created in young pulsars soon after the supernova
explosion or in binaries where the neutron star spins up (which also provides
extra mass to trigger the QN).

If the propeller works with the extreme accretion rate of
$\dot{m}= 10^{28}$~g~s$^{-1}$, then the rate of particle number is
$\dot{N}_{\rm prop.}=\dot{m}/(Z m_{\rm p})\sim 10^{52}/Z$~s$^{-1}$.
That is, after one crossing,
where a particle is boosted by a factor $2\Gamma_{\rm QN}^2$, the
energy in time unit used to accelerate the propeller wind is
$\dot{N}2\Gamma_{\rm prop.}\, m_{\rm p}c^2$; only the
particles with $\Gamma_{\rm prop}\ge \Gamma_{\rm QN}$ will
be accelerated by the QN shock.
This indicates that particle acceleration consumes the energy of the shock wave
 in $\sim 1$ ms, and the shock dies
out\footnote{Some of the ejecta may fall back allowing for disk and 
later planet
formation around the newly born QS \citep{keranen03}.}. We expect
the propeller to function efficiently in its early
stages with most particles acquiring $\Gamma_{\rm prop} \ge \Gamma_{\rm QN}$.
In reality
this timescale is longer because the propeller wind particles cannot 
propagate straightforwardly to the shock (due to magnetic field
and turbulence), their
energies must have spread out and the most energetic particles reach the shock
first; nevertheless, it is short compared to the timescale of the
expansion phase.
It also means that a tiny amount of the bulk of the propeller wind will be
accelerated to energies of the order of $10^{15}$~eV and at most up
to $10^{18}$~eV for those few particles that managed to get one more
kick in the process. Here again,
given the many similarities to the calculations and approach  
presented in \cite{achterberg01},
we expect the boosted propeller wind to acquire
a power law spectrum, ${\rm d}N/{\rm d}E\sim E^{-s}$ with
$s\simeq 3.2-3.3$.

Using the Milky Way dimensions with a radius of 15~kpc and a scale height of
$\sim$ 1~kpc, we obtain the number density of galactic cosmic rays for one QN
to be of the order of $n_{\rm MW}\sim 4\times 10^{-18}\, {\rm cm}^{-3}/Z\,
(10^{15}\, {\rm eV}/E)(E_{\rm QN}/10^{53}\,{\rm erg})$.
We assume a leaky box model \citep{munoz87,simpson88} for the 
galactic cosmic rays diffusing out
of the galaxy where we adopt the leakage timescale expressed
as  $T_{\rm d} = 3\times 10^7 \, {\rm yr}\, (E_{\rm GeV}/Z)^{-1/3}$ (see e.g. Webber 1998; Biermann et al., 2001 and Biermann \& 
Sigl, 2001).  The estimated
 time-averaged flux in our model is,
\begin{eqnarray}
J^{\rm QN}_{\rm 2}(E>10^{15}\, {\rm eV}) &\sim & \frac{c}{4\pi}
n_{\rm MW} T_{\rm d} \nu_{\rm QN} \nonumber \\
&\simeq & \frac{3\times10^{-10}}{Z^{2/3}}\, 
(\frac{10^{15}\, {\rm eV}}{E})^{4/3}
(\frac{\nu_{\rm QN}}{10^{-6}\,{\rm yr}^{-1}})
(\frac{\epsilon~E_{\rm QN}}{10^{52}\,{\rm erg}})
\,{\rm cm}^{-2}
{\rm s}^{-1} {\rm sr}^{-1}\, ,
\label{MWrays}
\end{eqnarray}
which is similar to the observed value of $J^{\rm
obs}(E>10^{15}\ {\rm eV})\sim
3\times10^{-10}$~cm$^{-2}$~s$^{-1}$~ster$^{-1}$ 
(\cite{bird95,candia03,haungs04}) 
as long as the hyper-Eddington accretion rates
can be accepted and if the accelerated particles are mainly protons. 
Notice that below the knee $J^{\rm QN}_{\rm 2}$ is too low to account for
the observed cosmic ray flux; the shock is
not energetic enough to accelerate more particles as to extend
the spectrum to lower energies.  
Therefore, the steeper spectrum above the knee
provided by the QN shock cannot dominate at energies below the knee. 
More important, the flux in~Eq.(14) is averaged over time, 
$t\gg 1/{\nu_{\rm QN}}$. Above the knee the cosmic ray 
diffusion time is shorter than the QN occurrence, 
$T_{\rm d}<\tau_{\rm QN}=1/{\nu_{\rm QN}}$. For example,
at the knee $T_{\rm d}\sim 3\times 10^5$ years for protons
and it is $T_{\rm d}\sim 3\times 10^4$ years at the ankle, while
the QN occurrence time is $\tau_{\rm QN}\sim 10^6$ years.
Therefore the propelled wind of the 
QN model can account for the galactic cosmic ray flux locally, i.e. in the 
vicinity and shortly after the QN event. Over the whole
galaxy and in timescales longer than millions of years, QNe 
can contribute on average a few percent of the galactic cosmic rays around 
the knee. Closer to the ankle the contribution is negligible due 
to the very short diffusion time $T_{\rm d}$. 
More accurate estimates
of the leakage times at high energies and QNe occurrence would be needed 
to make the above conclusions firm. In particular, the QN rate
carries substantial uncertainties; among
them the difficulty of  determining  the critical
density for the transition to quark matter, the 
 burden of not knowing which
 equation of state better describes the parent NS,
 and the lack of  a complete theory that can describe the QCD phase
diagram that would describe the path followed
by the NS during its transition to a quark star.
We note however that recent studies show that QN rates
can be as high $10^{-4}\ {\rm yr}^{-1}\ {\rm galaxy}^{-1}$
(Yasutake, Hashimoto, \& Eriguchi 2004).
which is above our fiducial value of
$10^{-6}\ {\rm yr}^{-1}\ {\rm galaxy}^{-1}$.

Finally, since the propelled wind consists of the crust material of the parent
NS, we expect this cosmic ray component to be rich
in heavier ions, up to iron. Their rigidity ($E/Z$) means that these
heavy nuclei would leak out of the galaxy 
at a lower rate than protons. It implies an increase of the heavy nuclei to hydrogen ratio
in the chemical composition over time for those cosmic rays
originating in QNe.

\subsection{Quark Nova as a relativistic particle injector}

In our model the propeller injects relativistic particles. 
As discussed above, some of this propelled 
material can be re-accelerated by the QN shock. The remaining
bulk will populate interstellar space. These particles with 
 Lorentz factors $\Gamma\sim 1-1000$ (or energies 1-1000~GeV for protons) 
could later be reaccelerated by galactic supernova shocks.
For the number of injected particles a simple estimate gives
$N_{\rm inj} \sim E_{\rm rot}/(10\, {\rm GeV})\sim
 10^{52}\, {\rm erg} / 10^{-2}\,{\rm erg}
  \sim 10^{54}$ 
that are accelerated to GeV energies by the propeller. Here 
$ E_{\rm rot}$ is  the maximum rotational energy of the QS.
Using Eq.~\ref{MWrays} this
implies $J \sim 10^{-3} \, {\rm cm}^{-2} {\rm s}^{-1} {\rm sr}^{-1}$,  
 which is a few percent of the observed value (see e.g. \cite{abe03}).
In this case the particle density in the galaxy can reach  
$n_{\rm MW} =  
4\times 10^{-13} {\rm cm}^{-3}$
from a single QN using up the QS rotational energy.
The spectrum of the population of these injected particles will be
shaped by later encounters and acceleration by galactic supernova shocks.
We note here as well that at most a few percent of the cosmic
rays below the knee can be of the QN origin.

The QN model predicts three cosmic ray components, two galactic
ones and an extragalactic one. This should in principle lead
into transitions or steps in the cosmic ray flux. The galactic
components will be subject to many SN shocks 
that will smoothen the discontinuity around the knee. Note
however that these signatures will be dwarfed by the 
much higher contribution from other galactic cosmic ray sources.
Having shown that QNe are candidate sources
of UHECRs above the ankle, we next discuss how observational
features can be explained within the QN model.

\section{UHECRs from QNe in future cosmic ray detectors}

With a rate of
$10^{-6}$ per galaxy per year and a galaxy density $n_{\rm
g}\simeq 0.02$~Mpc$^{-3}$ one should expect about $0.08$~QNe a
year within a $100$~Mpc sphere. 
This 100~Mpc region contains galactic and
intergalactic magnetic field regions to which GZK-energy
cosmic rays are subject to.
This may lead to  arrival time delay and therefore 
 clustering of events around the source. The
arrival timescales can spread from years to millions of years depending on the
strength and the configuration of the magnetic field as well
as the distance to the QN. Here we refer the interested reader 
to Sigl (2004a) for
a recent discussion on the effect of the
magnetic field on the ultra-high energy cosmic rays.
Given these numbers, there may be a tiny energy window with the particles
clustering within a few degrees towards the source
(e.g, Kronberg 1994a and 1994b).
This wishful possibility lead us to
consider such events in the future detectors
what we refer to as observational particle astronomy. Naturally,
such observations will be better guided if the QN can be detected
by other means as well (e.g. X-rays, gamma-rays).

The energy lost by the UHECRs as they propagate and interact with the
cosmic microwave background  
is transformed by cascading into secondary GeV-TeV 
photons (Protheroe\&Stanev 1996).
 This TeV-gamma-UHECR trace could
in principle be detected in the future (Catenese\&Weekes 1999)
and be used to test our model
 (see discussion in Akerlof et al. 2003 and references therein).
However photons and UHECRs may have very different 
arrival times that are not easily quantified.
This again calls for better understanding
of magnetic field effects before such connections
between the TeV photons and the related UHECRs can be made.
Also, since protons can be accelerated up to $10^{21}$ eV in our model,
significant neutrino fluxes (with energies above $10^{18}$ eV)
 can be generated (e.g.,  Waxman\&Bahcall 1999; Engel et al. 2001).
This is below the currently advertised threshold
of $5\times 10^{19}$ eV for EUSO and OWL
and most of the potential events will go undetected (Halzen\&Hooper 2002).
Nevertheless future neutrino detectors should be able
to signal any neutrino-UHECR trace with
the arrival direction of Ultra-High-Energy neutrinos as a good indication
of the QN location. The above mentioned traces are the subject
 of another study of QNe as sources of UHECRs.

\subsection{Multiple events and clustering of UHECRs}

Assuming that the cosmic rays above the ankle are accelerated in QNe, 
we obtain roughly $10^{44}$ particles above $10^{19}$~eV per QN.
The integrated flux F per unit area from one QN is 
\begin{equation}
F\simeq 10^{-10} (\frac{100\, {\rm Mpc} }{D_{\rm QN}})^2 \, {\rm cm}^{-2}\, ,
\end{equation} 
where the $D_{\rm QN}$ is the QN distance. This is the total flux 
integrated over time including the spread in arrival time induced 
by magnetic fields, which can vary from years to millions of 
years (Sigl 2004a). In a detector like Auger this implies (integrated 
over the year and aperture and assuming that the QN
is within the sky coverage) few hundred events per one QN at a distance
of 100~Mpc. In a detector like EUSO and OWL one can expect one to two orders
of magnitude more events per QN. If the time delay and thus the spread in 
arrival time is e.g. 1000 years, one could expect doublets and triplets
around a given QN within 10 year time in Auger, and equivalently dozens of
clustered events with EUSO and OWL. For completeness, if the average time
delay of cosmic rays above $10^{19}$~eV is around 1000~years, in
our model it would mean that UHECRs from
  roughly 100 QNe can presently be detected within 100~Mpc.

\subsection{Isotropy of UHECRs}

Unlike the arrival directions of UHECRs, 
the galaxy distribution is not isotropic within 100~Mpc 
distance. This may mean that
UHECR accelerators are not all within galaxies. 
It is natural to attribute  UHECR sources to galaxies
and the almost perfect  isotropy, sofar observed both below and above
the ankle (e.g. Torres\&Anchordoqui 2004), to 
 magnetic fields.
Despite the uncertainties on the magnetic field strength it has indeed been
shown that if our Local Supercluster
contains a large scale magnetic field it can provide  
sufficient bending to the cosmic ray trajectories 
(Sigl, Lemoine \& Biermann 1999; Farrar\&Piran 2000). 
Heavy ions like iron, if 
present in the pulsar wind, will have unique consequences.
Iron has a higher cutoff energy 
(\cite{puget76,stecker99}) than lighter nuclei thus allowing a longer path of propagation
in a tiny energy window allowing for more distance sources to contribute
to the flux.
Furthermore, being bent in galactic and intergalactic fields, heavy nuclei  
 would acquire a more isotropic distribution of arrival directions
than protons. 
Finally, heavy nuclei will naturally acquire higher energies than protons and contribute
to the extreme end of the UHECR flux in our model.

Of interest to the QN model, 
old population pulsars with large peculiar velocities\footnote{It is a well known fact that pulsars in our galaxy
have velocities much in excess of those ordinary stars
(Harrison, Lyne \& Anderson 1993).                                    
It is reported that their transverse
speeds range from $0$ to $\sim 1500$ km s$^{-1}$ and their
mean three-dimensional speed is $450\pm 90$ km s$^{-1}$
(Lyne \& Lorimer, 1994).} could 
cover tens of Mpc distances from their origins in a Hubble time.
Some of these run-away pulsars may undergo a QN explosion 
as a result of an increase
of their core densities following spin down or accretion from
the surrounding space. 
This would lean towards
a more isotropic distribution - the extent of isotropy in the
runaway pulsar population remains to be determined - of the
arrival directions of the UHECRs (the QN-shock boosted pulsar wind).
This favours weak magnetic field ($B < 10^{10}$ G) millisecond pulsars in our model, with the assumption 
that these old isolated pulsars still sustain their wind bubbles.
We note that these candidates are probably not born in the propeller 
regime following the QN explosion, given their weak magnetic
field and long periods (see \S~3).

\subsection{Injection spectrum and time-dependency}

For a given distant extragalactic QN source the more energetic
particles should arrive first since the associated time delay induced
by the intergalactic magnetic fields is short.
Therefore it would appear as if there is a monoenergetic flux
of particles from the source at one given time. Later,
the lower energy particles will arrive with larger time
delay (G. Sigl - private communication).

In the case of galactic cosmic rays, the propelled wind consists
of the crust material of the parent NS; the rigidity ($E/Z$)
implies that the heavy nuclei would leak out of the galaxy at a 
lower rate than that of protons. We thus expect an increase of the 
heavy-nuclei-to-hydrogen ratio in the chemical composition
over time scales of some ten thousand years for cosmic rays
originating in QNe. Note, however, that this is effect
is local and could have importance only if we detect a nearby
quark star. If the star has undergone a QN phase in the 
past, the local cosmic ray composition may reflect the time
passed since the QN explosion.

\section{Summary and conclusion}

The cosmic ray acceleration in the QN model consists of three different
components as illustrated in figure~1.
The first component is
due to the acceleration of the high energy particles in the
pulsar wind bubble by the QN shock.
We predict few hundred UHECR events above $10^{19}$ eV
for the Pierre Auger detector per a distant QN, while some thousands
are predicted for the proposed EUSO and OWL detectors.
Magnetic fields can lead into clustering of the predicted
events with timescales spread from years to millions of
years. 
 
 The second component stems from QNe 
with the compact remnant born in the propeller regime
(pulsars with high magnetic field and small period undergoing a QN
explosion immediately following a SN) ejecting  
 relativistic particles with  $\Gamma\sim 1-1000$. 
The particles with $\Gamma_{\rm prop} > \Gamma_{\rm QN}$ 
 will eventually interact with the QN shock, as to reach 
 roughly $10^{16}$~eV (protons) and $10^{18}$~eV (iron)
in one shock crossing. Given the energy of the QN shock, only a tiny 
amount of the propelled wind
particles can be accelerated, i.e. a maximum of $10^{50}$ particles
around the knee energy (see Eq.~\ref{MWrays}). 
 The propelled particles
that were not accelerated by the QN shock are injected
into the galactic space and will eventually become accelerated
by the supernova shocks in the galaxy. QNe as possible 
 UHECR sources seem to account for the observed extragalactic flux 
 and can contribute partially (a few percent) to the galactic cosmic
rays.
We conclude by stating that despite the fact that there is no guarantee 
that QNe occur in nature, the model possesses  
 features that can be tested in future cosmic ray detectors.

\begin{acknowledgements}
We thank the anonymous referee for the constructive comments that
 challenged our cosmic ray model and that helped us improve it.
PK and RO express their gratitude to the Department
of Physics at the University of Jyv\"askyl\"a, the Helsinki University
Observatory and the Science Institute of the University of Iceland for
hospitality. PK acknowledges the hospitality
of the University of Calgary  in the 
final stages of this work. The research of RO is supported by an 
operating grant from the Natural Science and Engineering Research 
Council of Canada (NSERC), as well as the
Alberta Ingenuity Fund (AIF). JM is supported by the Academy of Finland 
under the contracts no.\ 104915 and 107293, and PK under the contract
no.\ 106570. PK is also supported by the European Union
Regional Development Fund.
 
\end{acknowledgements}

\clearpage

\begin{figure}
\includegraphics[width=0.8\textwidth, angle=-90]{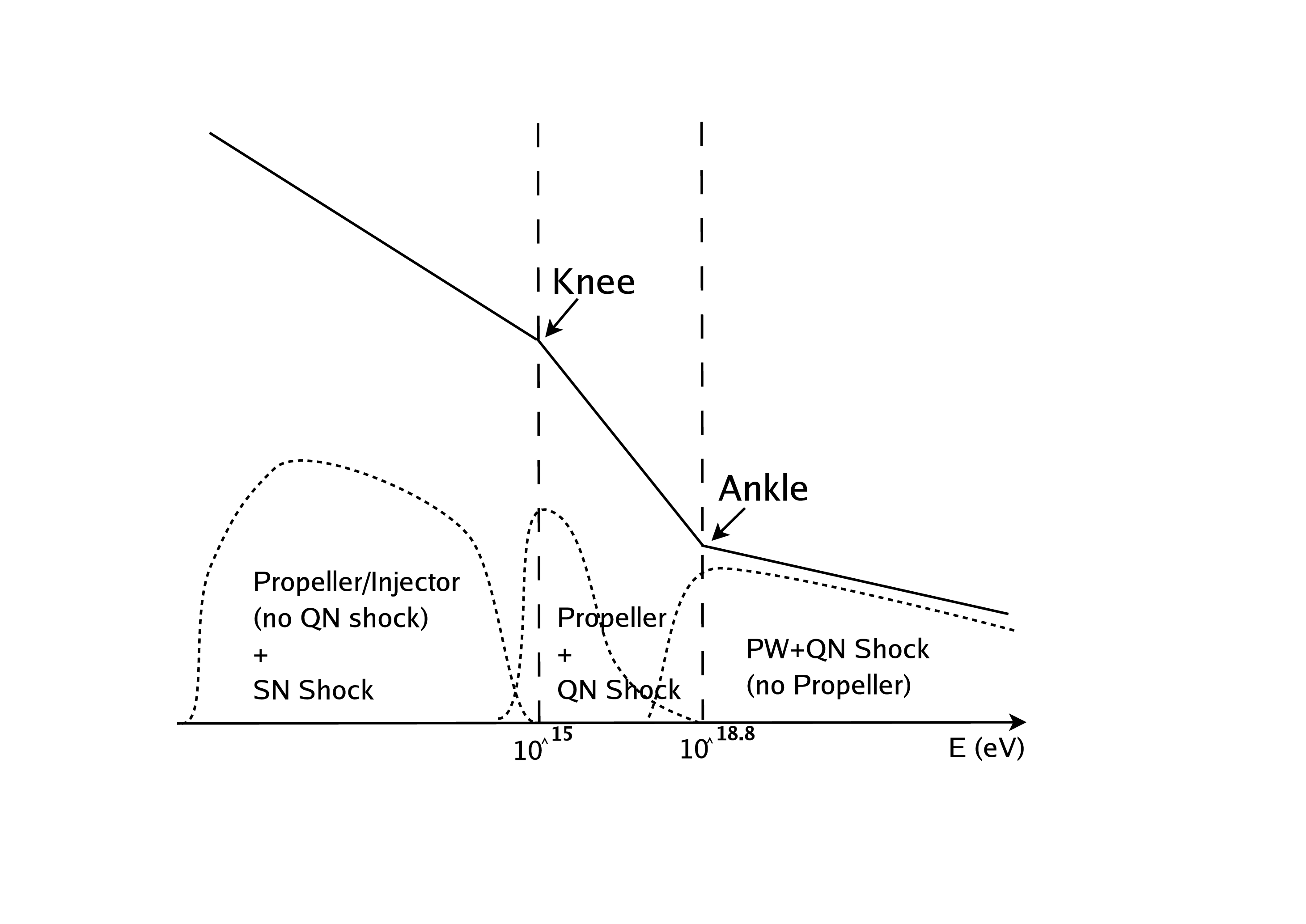}
\caption{Cosmic ray components in our model. 
The solid line illustrates the observed cosmic ray flux 
while the dotted line illustrates the contributions
from the QN model. Above the
ankle are the pulsar wind particles accelerated by the QN shock.
 Below the ankle and above the
knee are the particles first propelled
by the QN compact remnant and then boosted by the QN shock.
Only the most relativistic propelled particles will interact with the shock
which gets shut-off after a few milliseconds. Following this
latter phase the QN compact remnant acts as an injector
of relativistic particles which later would get accelerated by galactic
SN shocks to form the component with energies below the knee.
 Note that QNe as possible
 UHECR sources seem to account for the observed extragalactic flux (above
the ankle) and can contribute partially (a few percent) to the
 galactic cosmic rays (below and around the knee).
We note however that the observed extragalactic flux (above the ankle)
awaits experiments such as Auger, EUSO, and OWL
to resolve current disagreement between AGASA and HiRes.
\label{fig1}
}
\end{figure}


\clearpage

\begin{thebibliography}{}

\bibitem[Abe et al.(2003)]{abe03} Abe, K., et al. 2003, Phys. Lett. B564, 8 

\bibitem[Abu-Zayyad, T. et al.(2002)]{abuzayad02} Abu-Zayyad, T. et al. 2002 [astro-ph/0208301]

\bibitem[Achterberg et al.(2001)]{achterberg01} Achterberg, A., Gallant, Y. A., Kirk, J. G., \& Guthmann, A. W. 2001, MNRAS, 328, 
393

\bibitem[Akerlof et al.(2003)]{akerlof03} Akerlof et al. 2003, ApJ, 586, 1232

\bibitem[Alcock, Farhi, \& Olinto(1986)]{alcock86} Alcock, C., Farhi, E., \& Olinto, A. 1986, ApJ, 310, 261

\bibitem[Arons(2003)]{arons03} Arons, J. 2003, ApJ, 589, 871 

\bibitem[Backer et al.(1982)]{backer82} Backer, D. C. et al., 1982, Nature 300, 615

\bibitem[Bahcall \& Waxman(2003)]{bahcal03} Bahcall, J. N., \& Waxman, E. 2003, Phys. Lett. B, 556, 1

\bibitem[Berezinskii\&Grigor\'eva(1988)]{berezinskii88} 
 Berezinskii, V. S.; Grigor\'eva, S. I.
Astronomy and Astrophysics (ISSN 0004-6361), vol. 199, no. 1-2,
June 1988, p. 1-12

\bibitem[Bertone, Isola, \& Lemoine(2002)]{bertone02} Bertone, G., Isola, C., Lemoine, M., \& Sigl, G. 2002, Phys. Rev. D. 66, 
103003

\bibitem[Biermann\&Stritmatter(1987)]{biermann87} Biermann, P. L., \& Strittmatter, P. A. 1987, ApJ, 322, 643

\bibitem[Biermann(1998)]{biermann98} Biermann, P. L.: in: Workshop on Observing
Giant Cosmic ray Showers from Space, ed. J. F. Krizmanic, J. F.
Ormes and R. E. Streitmatter, American Institute of Physics, New
York (1998), p. 22

\bibitem[Biermann et al.(2001)]{biermann01} Biermann, P. L., Langer, N., Seo, E.-S., \& Stanev, T. 2001, 
A\&A 369, 269

\bibitem[Bierman \& Sigl(2001)]{biermansigl01} Biermann, P. L., \& Sigl, G. 2001, Lect. Notes Phys. 576, 1,
(Springer, Heidelberg) (astro-ph/0202425)

\bibitem[Bird et al.(1995)]{bird95} Bird, D. J.,  et al. 1995, ApJ, 441, 144

\bibitem[Blandford \& McKee(1976)]{blandford76} Blandford, R. D., \& McKee, C. F. 1976, Phys. Fluid. 19, 1130

\bibitem[Blasi(2000)]{blasi00} Blasi, P., Epstein, R. I., \& and Olinto, A. V. 2000, ApJ 533 L123

\bibitem[Blasi(2003)]{blasi03} Blasi, P. astro-ph/0304206

\bibitem[Bombaci \& Data(2000)]{bombaci00} Bombaci, I. \& Datta, B. 2000, ApJ, 530, L69

\bibitem[Candia(2003)]{candia03} Candia, J., Mollerach, S., \& Roulet, E., JCAP 05(2003)003

\bibitem[Catanese\&Weekes(1999)]{catanese99} Catanese, M. \& Weekes, T. C. 1999, Publ. Astron. Soc. of Pacific,
Vol. 111, issue 764, 1193

\bibitem[Cheng \& Dai(1996)]{cheng96} Cheng, K. S., \& Dai, Z. G. 1996, Phys. Rev. Lett., 77, 1210

\bibitem[Chevalier(1989)]{cheval89} Chevalier, R. N. 1989, ApJ, 346, 847

\bibitem[Dai \& Peng(1995)]{dai95} Dai, Z. G., Peng, Q. H., \& Lu, T. 1995, ApJ, 440, 516

\bibitem[Engel, Seckel, \& Stanev(2001)]{engel01} Engel, R., Seckel, D., \& Stanev, T. 2001, Phys. Rev. D., 64, 093010

\bibitem[Farrar \& Piran(2000)]{farrar00} Farrar, G. R., \& Piran, T. 2000, Phys. Rev. Lett. 84, 3527 

\bibitem[Fendt \& Ouyed(2004)]{fendt04} Fendt, Ch. \& Ouyed, R. 2004, ApJ, 608, 378 

\bibitem[Frank, King, \& Raine(1992)]{frank92} Frank, J., King, A., \& Raine, D., Accretion Power in Astrophysics,  Cambridge 
Univ. Press, Cambridge (1992)

\bibitem[Gallant \& Arons(1994)]{gallant94} Gallant, Y. A., \& Arons, J. ApJ, 435 (1994), 230

\bibitem[Gallant \& Achterberg(1999)]{gallant99} Gallant, Y. A., \&  Achterberg, A. 1999, MNRAS, 305, L6

\bibitem[Garcia-Munoz et al.(1987)]{munoz87} Garcia-Munoz, M.,
Simpson, J.~A., Guzik, T.~G., Wefel, J.~P., \& Margolis, S.~H.\ 1987,
\apjs, 64, 269

\bibitem[Goldreich \& Julian(1969)]{gj69} Goldreich, P., \& Julian, W. H. 1969,   ApJ, 157, 869

\bibitem[Greisen(1966)]{greisen66} Greisen, K. 1966, Phys. Rev. Lett. 16,  748

\bibitem[Halzen\&Hooper(2002)]{halzen02} Halzen, F. \& Hooper, D. 2002 Rep. Prog. Phys. 65 1025-1078

\bibitem[Harrison, Lyne, \& Anderson(1993)]{harrison93} Harrison, P. A., Lyne, A. G., \& Anderson, B. 1993, MNRAS, 261, 113 

\bibitem[Haungs(2004)]{haungs04} Haungs, A. et al., 2004, Acta Phys. 
Pol. B 35, 331

\bibitem[Hoshino(1992)]{hoshino92} Hoshino, M. et al. 1992, ApJ, 390, 454

\bibitem[Illarionov \& Sunyaev(1975)]{illarionov75} Illarionov, A. F., \& Sunyaev, R. A. 1975,  A\&A, 39,  185

\bibitem[Ker\"anen \& Ouyed(2003)]{keranen03} Ker\"anen, P., \& Ouyed, R. 2003, A\&A, 407, L51  

\bibitem[Ker\"anen, Ouyed \& Jaikumar (2005)]{keranen05} Ker\"anen, P., 
Ouyed, R., \& Jaikumar, P., 2005, ApJ, 618, 485 

\bibitem[Kronberg(1994a)]{kronberga} Kronberg, P. P. 1994a, Rep. Prog. Phys. 57, 325

\bibitem[Kronberg(1994b)]{kronbergb} Kronberg, P. P. 1994b, Nature (London) 370, 179

\bibitem[Kundt(1990)]{kundt90} Kundt, W.: in:  Neutron stars and their birth events, ed.  W. Kundt, Kluwer Academic Publishers, 
Dordrecht (1990), p. 1

\bibitem[Lawrence, Reid, \& Watson(1991)]{lawrence91} Lawrence, M. A., Reid, R. J. O., \& Watson, A. A. 1991, J. Phys. G., 17, 733

\bibitem[Lyne, \& Lorimer(1994)]{lyne94} Lyne, A., \& Lorimer, D. R. 1994, Nature, 369, 127

\bibitem[Manchester \& Taylor(1977)]{manchester77} Manchester, R. N., \& Taylor, J. H., Pulsars, Freeman, San Francisco (1977)

\bibitem[Menou et al.(1999)]{menou99} Menou, K., et al. 1999, ApJ, 520, 276

\bibitem[Nagano \& Watson(2000)]{nagano} Nagano, M \& Watson, A. A. 2000, Rev. Mod. Phys. 72, 689 

\bibitem[Olinto(2004)]{olinto04} Olinto, A. V. 2004, astro-ph/0404114

\bibitem[Ostrowski(2002)]{ostrowski02} Ostrowski, M. 2002, Astropart. Phys. 18,  229

\bibitem[Ouyed, Dey \& Dey(2002)]{ouyed02} Ouyed, R., Dey, J., \& Dey, M. 2002,  A\&A, 390, L39

\bibitem[Protheroe\& Stanev(1996)]{protheroe96} Protheroe, R. J., \& Stanev, T. 1996, Phys. Rev. Lett., 77, 3708

\bibitem[Puget, Stecker, \& Bredekamp(1976)]{puget76} Puget, J. L., Stecker, F. W., \& Bredekamp, J. H. 1976, ApJ, 205, 638

\bibitem[Shvartsman(1970)]{shvartsman70} Shvartsman, V. F. 1970, Radiofizika 13, 1852

\bibitem[Shapiro \& Teukolsky(1983)]{shapiro83} Shapiro, S. L., \& Teukolsky, S. A., Black Holes, White Dwarfs, and Neutron Stars, 
Wiley, New York
(1983)

\bibitem[Sigl et al.(1994)]{sigl94} Sigl, G., Schramm, D. N., \& Bhattacharjee, P. 1994, Astropart. Phys. 2, 401

\bibitem[Sigl,Lemoine\& Bierman(1999)]{sigl99} Sigl, G., Lemoine, M., \& Biermann, P. 1999, Astr. part. phys., 10, 141 

\bibitem[Sigl(2003)]{sigl03} Sigl, G. 2003, Ann. Phys., 303, 117

\bibitem[Sigl(2004a)]{sigl04b} Sigl, G. 2004, astro-ph/0404074 (2004a)

\bibitem[Sigl(2004b)]{sigl04b} Sigl, G. 2004, JCAP, 8, 012 (2004b) 

\bibitem[Simpson, \& Garcia-Munoz(1988)]{simpson88} Simpson, J. A., \& Garcia-Munoz, M. 1998, Space Sci. Rev. 46, 205

\bibitem[Stanev, Seckel, \& Engel(2003)]{stanev03} Stanev, T., Seckel, D., \& Engel, R. 2003, Phys. Rev. D, 68, 103004

\bibitem[Stanev(2004)]{stanev04} Stanev, T. 2004, astro-ph/0411113 

\bibitem[Stecker, \& Salomon(1999)]{stecker99} Stecker, F. W., \& Salomon, M. H. 1999, ApJ, 512, 521

\bibitem[Takeda et al.(1998)]{takeda98} Takeda, M. et al., 1998, Phys. Rev. Lett. 81, 1163

\bibitem[Takeda et al.(1999)]{takeda99} Takeda, M. et al. 1999, ApJ, 522, 225

\bibitem[Torres \& Anchordoqui(2004)]{torres} Torres, D. F. \& Anchordoqui 2004 [astro-ph/0402371] 

\bibitem[Venkatesan et al.(1997)]{venkatesan97} Venkatesan, A., Miller, M. C., \& Olinto, A. V. 1997, ApJ, 484, 323

\bibitem[Waxman\&Bahcall(1999)]{waxman99} Waxman, E., \& Bahcall, J. 1999, Phys. Rev. D. 59, 023002

\bibitem[Webber et al.(1998)]{weber98} Webber, W. R.,  et al., 1998, ApJ, 508, 940

\bibitem[www.auger.org(2004)]{auger} www.auger.org

\bibitem[Xu, Zhang, \& Qiao(2001)]{xu01} Xu, R. X., Zhang, B., \& Qiao, G. J. 2001, Astropart. Phys., 15, 101

\bibitem[Xu(2003)]{xu03} Xu, R. 2003, in Stellar astrophysics - a tribute to Helmut A. Abt. Sixth Pacific Rim Conference, Xi'an, 
China, 11-17 July 2002. Edited by K. S. Cheng, K. C. Leung and T. P. Li (Astrophysics and Space Science Library, Vol. 298, 
Dordrecht: Kluwer Academic Publishers, ISBN 1-4020-1683-2, 2003, p. 73 - 81) 

\bibitem[Yasutake, Hashimoto, \& Eriguchi(2004)]{yasutake04} Yasutake,
N. Hashimoto, M.-A., Eriguchi, Y. 2004 [astro-ph/0411434]

\bibitem[Zatsepin \& Kuzmin(1966)]{zatsepin66} Zatsepin, G. T., \& Kuzmin, V. A. 1966,  Pis'ma Zh. Eksp. Teor. Fiz. 4, 114 ({\it 
JETP. Let.} 4 (1966), 78)

\bibitem[Zhang, Xu \& Qiao(2000)]{zhang00} Zhang, B., Xu, R. X., \& , Qiao, G. J. 2000, ApJ, 545, L127

\end{thebibliography}
\end{document}